\documentclass[runningheads]{llncs}
\usepackage{graphicx}
\begin{document}
\title{Forgotten @ Scale: A Practical Solution for Implementing the Right To Be Forgotten in Large-Scale Systems}
\titlerunning{Forgotten @ Scale}
%
\author{Abigail Goldsteen\inst{1} \and
Tomer Douek\inst{2} \and
Yaniv Cohen\inst{2} \and
Igor Gokhman\inst{1} \and
Ofir Keren-Ackerman\inst{2} \and
Gadi Katsovich\inst{3} \and
Grisha Weintraub\inst{3} \and
Doron Ben-Ari\inst{2}}
\authorrunning{A. Goldsteen et al.}
%
\institute{IBM Research \email{abigailt,igorgok@il.ibm.com} \and 
	IBM \email{tom,yanivc,ofirke,doronb@il.ibm.com} \and 
	IBM \email{Gadi.Katsovich,Grisha.Weintraub@ibm.com}}
\maketitle              
\begin{abstract}
The European General Data Protection Regulation asserts data subjects' right to be forgotten, i.e., their right to request that all their personal data be deleted from an organizations' data stores. However, fulfilling such requests in large-scale systems is technically challenging. It requires that organizations keep track of all locations in which an individual's data is stored, be able to access and delete it in a reasonable time frame, and be able to prove that all such data was in fact deleted. In addition, organizations must cope with complexities such as multiple, distributed, and continuously evolving systems of record, complex data retention policies and deletion approval workflows. We present a first design pattern and practical implementation of the right to be forgotten on a large scale in Big Data and cloud environments. 

\end{abstract}
\section{Introduction}
The European General Data Protection Regulation (GDPR)\footnote{https://www.eugdpr.org/}, which went into effect in May 2018, replacing the previous Data Protection Directive from 1995. According to the GDPR, \textit{personal data} is defined as ``any information relating to an identified or identifiable natural person", called the \textit{data subject}. The regulation distinguishes between two key roles: the \textit{data controller} and the \textit{data processor}. The \textit{data controller} is the person, public authority, agency or any other body (e.g., company) which determines the purposes and means of the processing of personal data. The \textit{data processor} is a person, public authority, agency or any other body that processes personal data on behalf of the controller.

Article 17 of this regulation defines the ``right to erasure", also referred to as the ``right to be forgotten": a data subject, under certain conditions, shall have the right to obtain from the data controller the erasure of personal data concerning him or her without undue delay. The Article additionally states that the right to erasure does not apply if further processing of the data is necessary to fulfill a legal obligation, for purposes that are in the public interest, or for the defense of legal claims. Such a situation may arise if another law or regulation prevents the deletion of certain data. For example, in the financial industry, a European regulation called MiFID II\footnote{https://ec.europa.eu/info/law/markets-financial-instruments-mifid-ii-directive-2014-65-eu\_en} requires that transaction details be kept for at least five years.

According to Article 19 of the GDPR, the controller is required to communicate any such erasure request to each recipient to whom the personal data have been disclosed, so that they may erase it as well. This means that the right to erasure also applies to data processors. 

Being able to fulfill such erasure requests is technically very challenging. It requires organizations to keep track of all locations in which an individual's personal data is stored, including derived data (data that was computed based on raw data collected from the subject). For each such data they must determine whether it should be deleted or one of the exemptions set out in the regulation apply. In addition, organizations must be able to find and delete all relevant data within a reasonable time frame, usually one to two months, and be able to prove that all data was in fact deleted. If part of the data cannot be deleted for one of the above-mentioned reasons, this must also be documented and communicated to the data subject when replying to their request. 

Existing solutions mostly tackle the challenge of the deletion process itself, verifying that the data is securely deleted and cannot be restored, wihtout dealing with actually identifying and finding which data to delete. Another set of solutions relies on simple string matching or regular expressions to locate the data of an individual. For example, for a person named John Doe with the national identifier 12345 and email address j@email.com, these solutions simply search within data stores for those (or similar) strings. Clearly, not all personal data can be so easily located, as it may have been pseudonymized and no longer associated with the person's original identifiers. 

While the above examples address various aspects of the deletion problem, no solution that we know of tackles the end-to-end challenge of the data deletion lifecycle in highly complex and distributed systems. This includes locating data across multiple systems and instances, retention policies that may override the right to erasure, evidence retention, and audit logging. This end-to-end lifecycle becomes a critical problem in Big Data environments where systems are geographically dispersed, constantly evolving and are typically owned and operated by different groups or business units. Moreover, these systems serve different purposes and thus may require different approaches for implementing deletion procedures.

In this paper, we present a novel, practical solution to implement the right to be forgotten in large-scale Big Data, cloud environments. We introduce a reusable design pattern for recording and registering erasure requests, launching deletion  processes across multiple systems, tracking their progress until completion, and issuing a report including details on what data was deleted. The framework is described in detail in Section \ref{Solution} of this paper. We then present a practical implementation of the framework in a real IBM product. The details of this implementation are described in Section \ref{IBM}, which also includes insights on the challenges encountered and the design decisions that were made. Finally, we conclude our work and outline future work directions in Section \ref{Conclusion}.

\section{Related work}
\label{Related work}
One area that has been pursued in connection with the right to erasure is the issue of secure deletion, i.e., making sure the deleted data cannot be restored. Works in this area include TrueErase \cite{TrueErase}, FullPP \cite{fullpp} and Multi-User Secure Deletion \cite{cloud}. The Ephemerizer \cite{eph} uses encryption to manage data access and Vanish \cite{vanish} uses Distributed Hash Tables to automatically expire data after a defined time period. Commercial solutions include Eraser\footnote{https://eraser.heidi.ie/} and Certus Erasure\footnote{https://www.certus.software/en/}. Velupillai et al. \cite{shredders} provide an evaluation of different file shredding software programs for the Windows operating system.

Other facets of the GDPR have also been widely researched and implemented. A multitude of companies offer tools for personal data discovery and classification, including Microsoft\textregistered data discovery tools\footnote{https://docs.microsoft.com/en-us/azure/security/how-to-discover-classify-personal-data-azure} and the IBM\textregistered\  StoredIQ Suite\footnote{https://www.ibm.com/us-en/marketplace/ibm-storediq-suite}. Academic works on this topic include a tool for classifying enterprise data using Semantic Web technologies \cite{enterprise}, as well as many works on document classification \cite{classification,neural}. Recent solutions such as MinerEye\footnote{https://minereye.com/} use artificial intelligence to identify and classify sensitive data.

Consent management solutions are quite abundant, both in the literature and in the commercial arena. Early works include the Encore project \cite{encore} and Smart Notices \cite{CC}. More recent works include DPCM \cite{dpcm}, SPECIAL \cite{scalable} and Semantic Based Consent Model \cite{model}. Commercial solutions include Symphonic\footnote{http://www.symphonicsoft.com/solution/}, OneTrust\textregistered\footnote{https://onetrust.com/products/gdpr-compliance/}, and more. The OneTrust solution also includes a portal that data subjects can use to submit their erasure requests. The portal notifies process owners of the requests and automatically communicates the responses to data subjects, alongside generating compliance reports. However, it does not cover the process of actual deletion from the organizations' systems and data stores.

Another set of tools is designed to monitor all data accesses to ensure they are compliant with policies and to prevent data breaches. Such tools include IBM Security Guardium\textregistered\footnote{https://www.ibm.com/security/data-security/guardium} and Imperva\textregistered\ SecureSphere\textregistered\footnote{https://www.imperva.com/products/data-security/data-protection/}. Enterprise cloud providers have also started to implement activity monitoring and audit capabilities, e.g., Amazon\textregistered\ AWS CloudTrail\footnote{https://aws.amazon.com/cloudtrail/}, and IBM Cloud Activity tracker\footnote{https://www.ibm.com/cloud/activity-tracker}. Forsberg in his thesis \cite{central} suggested the use of a System Information and Event Management (SIEM) or log management solution to monitor all access to files containing regulated data.

Another approach is to use information-flow techniques to track the flow of personal or sensitive information within applications and ensure it is properly used. The Decentralized Information Flow Control model \cite{flow}, RIFLE \cite{rifle} and TaintDroid \cite{taint} and its many successors fall under this category.

Additional work \cite{google} includes a description of how Google\textregistered\ and other search engines implemented the right to be forgotten in the aftermath of the famous Google Spain verdict in 2014. However, this report only describes what the right to erasure looks like from the end-user's perspective, e.g., the portal for submitting requests, what questions need to be answered, etc. It does not provide insight about how this was implemented by Google behind the scenes. Moreover, this example only covers the removal of specific URLs from search results, not the removal of all personal data.

Malle et al. \cite{ml} tackled the issue of the right to be forgotten from a different perspective, seeking to determine whether existing machine learning techniques could be effective when applied on anonymized data, thus exempting it from the need to be erased. They tested four classifiers, running them on an initial dataset and then on purturbed versions of it, and recording the impact on the quality of the classification result. As expected, the accuracy declined significantly in all tested algorithms, leading the authors to the conclusion that anonymizing data before applying classification is not a viable solution for most use cases.  ROBUST-STREAMING \cite{sum} is a resilient streaming algorithm aimed at solving the problem of summarizing a dynamic data stream while enabling users to restrict the service provider from using their data at any time.

O'Hara \cite{semantic} proposed a Semantic Web based approach for implementing the right to be forgotten. However, that proposal is more of a high-level discussion than a concrete technical solution. Sarkar et al. \cite{erasure} present the research challenges in facilitating the erasure of data as per the right to erasure and propose some technical solutions for those challenges. The two main challenges they identified were identification of data replication and designing distributed data erasure algorithms. As possible solutions they propose exploiting operating system capabilities, tainting methods, encryption-based erasure techniques, and "eventual erasure" for hard-to-access locations. However, they did not present an actual implementation.

The main research gap remains how to locate and track all places where a person's data is stored (including derived data), determine which ones can and should be deleted, delete the data across many, distributed systems, track the deletion process and be able to report on it in a timely manner. This is a very complex problem, encompassing many challenges, some of which are solved by our framework and some remain future work. 

\section{The data deletion framework}
\label{Solution}
In this section, we present a design pattern for managing and executing data subject erasure requests. This pattern was designed to be as generic as possible, to support different types of IT systems and easily add new ones if needed. It was also designed to support erasure requests not only for a specific data subject but also for entire applications or lines of business, thus making it reusable for additional use cases such as fulfilling contractual obligations between companies transferring data from one to the other for business purposes. 

\begin{figure*}[htbp]
	\centerline{\includegraphics[width=12cm]{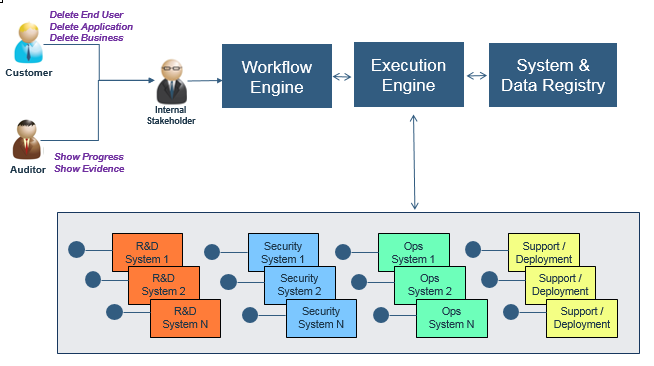}}
	\caption{High-level architecture of the data deletion solution}
	\label{arch}
\end{figure*}

The framework has three main building blocks: 1. A \textbf{System and Data Registry} to maintain the information about where data is stored and how to delete it. 2. A \textbf{Workflow Engine} in which erasure requests are created, tracked, and documented. 3. An \textbf{Execution Engine} for managing and triggering deletion jobs. The overall architecture of the solution can be seen in Figure \ref{arch}. On the top are the three main components of the solution, and on the bottom are the individual systems where deletion is performed. In the following sub-sections, we provide more details on each of these components.

\subsection{System and Data Registry}
\label{registry}
The System and Data Registry maps the different data items to the systems in which they are stored, and holds additional operational information to support automatic and scalable deletion execution. The registry contains definitions of the various data usage purposes, for example, research, authentication, malware detection, etc. It also includes the types of personal data collected and stored in the systems, for example IP address, geo-location, device profile, and so forth. The different system types are also defined. Each system type can be linked to one or more data types that are stored in it and one or more data usage purposes.

Finally, all production systems used to store and process personal data are registeredy. Each such system is an instantiation of one of the system types that were previously defined. Thus, several identical systems instantiated in different data centers, spanning different geographies, are supported. For each system, its name, geographic location, data center identifier, and additional details such as the system owner and business owner are stored. For each system, relevant information on how to trigger the actual deletion is also documented. 

Each system can also be linked to a retention policy that states for how long data should be stored on the system. For example, a policy could indicate that data should be stored for 30 days, 90 days, or one year, or for the duration of a certain contract between two parties. These retention policies can be used to implement an automatic deletion tool for "expired data", whose retention period has passed. Retention policies can also be linked to a specific purpose, e.g., for systems with more than one purpose, data used for different purposes could have different retention periods.

\subsection{Workflow Engine}
\label{workflow}
The second building block is a Workflow Engine, used to define the deletion lifecycle, track deletion progress, retain evidence, and log information for auditing purposes. 
To initiate the process a deletion request is created and logged. Requests can be issued manually, by a person, or automatically via code invocation (e.g., triggering the creation of a deletion request via an API). Such requests must indicate the user to be deleted. User identification can be done by various mechanisms such as a unique user ID, an email address, etc.

Depending on how the task workflow is configured, it may require review or approval by one or more people before it goes to the execution stage. It may also involve a process for checking whether the data should not be deleted due to one of the exemptions set out in Article 17 of the GDPR, e.g., further processing of the data is required for exercising the right of freedom of expression and information; compliance with a legal obligation; archiving purposes in the public interest, scientific or historical research purposes; or for the establishment, exercise or defence of legal claims. In the case of health related data, it may also be retained for public health reasons.

The request then proceeds to the execution phase. As the deletion process is expected to span multiple systems, a sub-task is created to track the progress of the deletion on each individual system from which data needs to be deleted. In some systems a deletion request may not complete instantly and may take hours or even days to process. When each system-level deletion request terminates, evidence that the system performed the deletion is attached to the sub-task, demonstrating either successful execution of the request, failure to delete due to an error, or confirmation that no data matching the search criteria was found. Each such sub-task is then marked as completed. Once all sub-tasks are complete, the overall master task can be closed, either automatically or following a human review. 

In addition to managing this workflow, the Workflow Engine also serves as an audit trail retention system, as each system sub-task is augmented with execution evidence. The evidence collected during the deletion process can be supplied to the data subject or to an auditor, to show that the deletion steps were executed successfully. The Workflow Engine also allows users to view the deletion tasks' status and register to receive notifications upon events such as task creation, approval, or termination. 

\subsection{Execution Engine}
\label{execution}
The Execution Engine serves as a data deletion job launcher and tracker and is used to trigger the actual deletion tasks in an automatic manner. A job is run periodically to check for new erasure requests in the Workflow Engine that are ready for execution. For each such request, the job queries the registry to get the list of production systems from which data needs to be deleted, along with the deletion commands that need to be run for each system. Deletion commands are invoked using an API implemented by the system; the API can either be externalized using HTTP or invoked by running a jar file or script, optionally with certain parameters. 

Since different systems have different deletion logic and requirements, the ability of system owners to develop and own a proper deletion mechanism for their system is crucial. In addition, in typical cloud deployments, identical systems are often instantiated in different geographies. As a result, a deletion function developed for one system can be reused (or reused with small variations) across multiple regions.

For each relevant system, the Execution Engine first opens a sub-task in the Workflow Engine to represent the specific system-level deletion task. It then runs the relevant deletion commands and waits to receive acknowledgement from the system that the task has been completed.  Eventually, when the system-level deletion is complete, the system informs the Execution Engine of the completion and sends any relevant evidence. Such evidence may be a snippet from a log file, the response or return code of the command, and so forth. The Execution Engine then attaches the evidence to the appropriate sub-task and marks it as complete. 

\section{IBM use case}
\label{IBM}
IBM offers many different products and solutions to its clients, including middleware, software and cloud-based services in the areas of cloud computing, artificial intelligence, commerce, data and analytics, Internet of Things (IoT), mobile, and security. As a data processor, IBM may receive erasure requests from its clients, the data controllers, concerning certain data subjects and must comply with those requests.

One of IBM Security's products is a large scale Software-as-a-Service (SaaS) product, spanning multiple geographies and consisting of thousands of servers, hundreds of databases, implemented using multiple technologies. The remainder of this paper is focused on the solution put in place for this specific product. 

In this implementation, a Data Policy and Consent Manager (DPCM) \cite{dpcm} instance was used as the System and Data Registry, and for storing the data retention policies that may, in some cases, override the erasure request; Jira\textregistered\footnote{https://www.atlassian.com/software/jira} was used as the Workflow Engine; and a Jenkins\textregistered\footnote{https://jenkins.io/} instance, augmented with proprietary deletion management code, was used as the Execution Engine. The implementation of these components follows the conceptual framework described in Section \ref{Solution}. An additional Management component was also implemented to manage the specific deletion jobs for each target system. 

\begin{figure*}[htbp]
	\centerline{\includegraphics[width=12cm]{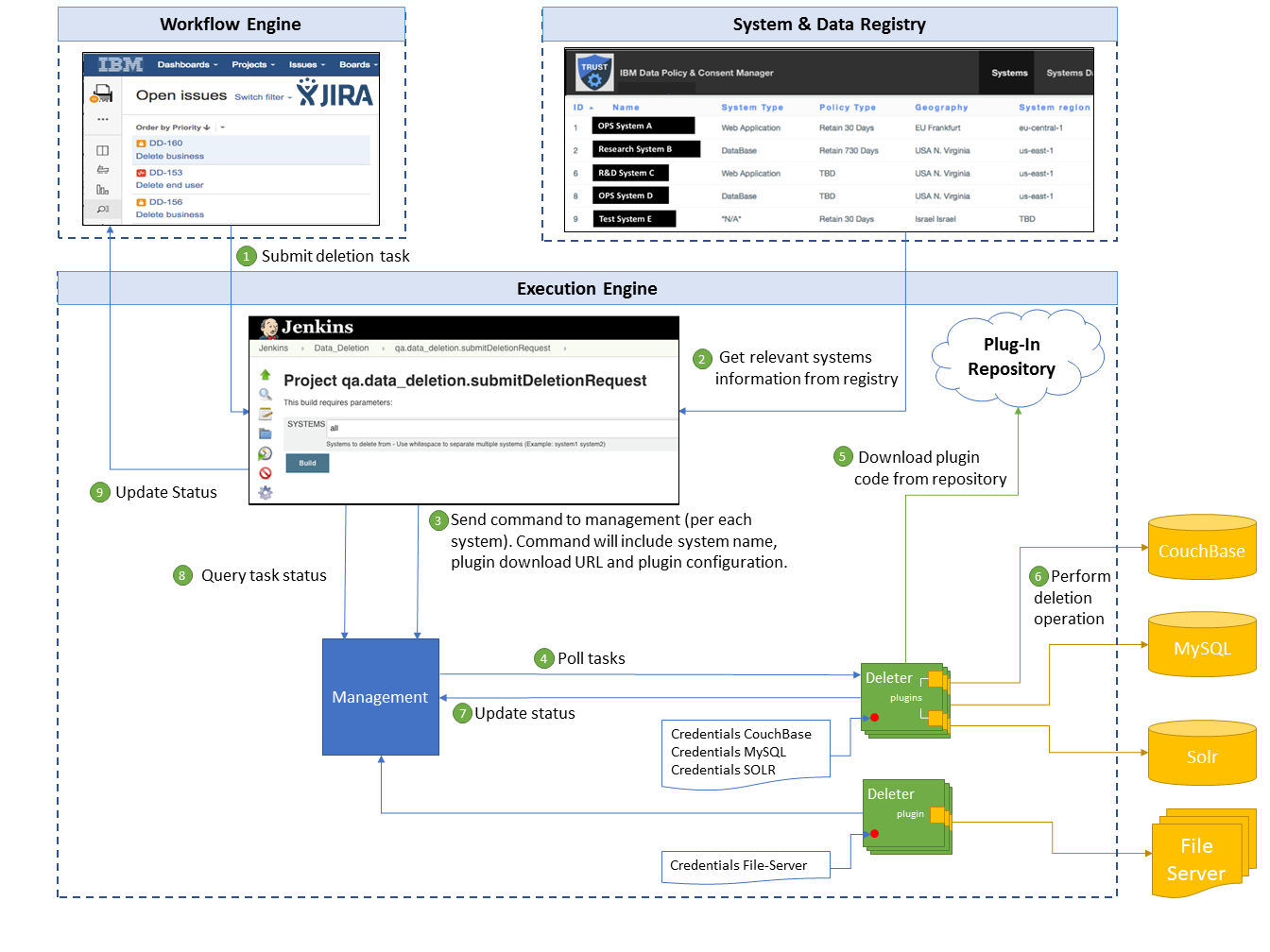}}
	\caption{IBM's data deletion flow}
	\label{details}
\end{figure*}

IBM implemented the system-specific deletion tools in a unified manner, using plugins that can be downloaded and executed on a target system. The plugins can be jar files or python modules containing code that implements the deletion logic. All plugins are hosted in a central repository to enable easy maintenance and updates of the code if necessary. A more detailed execution flow of the deletion process can be found in Figure \ref{details}. The following sub-sections elaborate on some aspects of the implementation. 

\subsection{DPCM component}
\label{DPCM}
The Data Policy and Consent Management tool \cite{dpcm} was designed to manage many aspects related to the policies governing the use of data in the enterprise. It includes tools for modeling and storing privacy policies and user consent; mechanisms for certifying the purpose(s) of an application; and a policy decision engine that handles the data access logic. The decision engine helps determine whether a data item can be accessed in a certain context. This engine processes all policies applicable to the data access, and as a result can either approve the access, deny the access or check whether the data subject(s) gave consent and approve/deny based on that. Another possible response coould be that the data must be obfuscated (masked or redacted in some way) before being used for the requested purpose. To support the right to erasure, only a subset of DPCM's capabilites are used. 

For each system, the relevant information on how to trigger the actual deletion is documented in DPCM. In this case, the information includes a URL to the relevant Management system, the name of the plugin to download, and the relevant plugin configuration for that target system. Thus, similar systems may reuse the same plugins but run them with different configurations.

DPCM's policies can be used to describe additional rules that apply to the data, and may, in some cases, override the deletion request. For example, a policy can state that transaction data must be stored for five years (as in the example given earlier). In this case, when calling DPCM's decision engine to request approval for deleting a person's past transactions, if the data's timestamp is less than five years old, the API will return a response of 'deny', stating that the decision was made based on the above policy. This response can then be returned to the Execution Engine to be logged with the deletion request.

\subsection{Jenkins component}
\label{Jenkins}
This component is based on the open source Jenkins automation server software. A Jenkins job is run periodically to check for new erasure requests (tickets) in Jira that are ready for execution. For each such ticket, it queries DPCM to get the list of systems from which data needs to be deleted, along with the deletion commands that need to be run for each one. These commands, in turn, are sent to a Management component that communicates with the individual systems. The deletion request includes additional information taken from the system record in DPCM, such as the plugin URL and plugin config, which are used by the Management component to trigger the deletion job (see Section \ref{Management}).

\subsection{Management component}
\label{Management}
The Management component is in charge of managing the specific deletion jobs for each target system. It also manages the access credentials for the systems it is resposible for. The Management component receives a deletion request from Jenkins for a specific user ID (or business ID if the request is to delete an entire business), along with the list of systems retrieved from DPCM.  

For each system, the Management component generates a job containing all relevant information to perform the deletion. Each system's deletion tool polls the Management system periodically to discover new deletion jobs that are waiting to be executed, and initiates the execution. Long deletion processes are broken into multiple execution steps, after each one of which the plugin reports its status as "in progress". 

Eventually, when the deletion is complete, the plugin informs the Management component that the task has been completed and sends along any relevant evidence. 

\subsection{Challenges in the Solution Implementation}
The first challenge we encountered was dealing with online, highly available systems. Since running delete operations on a database may significantly impact its performance, we had to plan the solution such that it would not degrade these systems' performance or availability. This was done by enabling the deletion plugins to delay execution of the deletion command if needed, for example if the current load on the system is above a certain threshold. 

The second challenge we dealt with was systems that perform intensive data loading operations. For example, a system may be populated by periodically running an Extract-Transform-Load (ETL) job that loads large quantities of data into the datastore. To prevent deletion jobs from interfering with the ETL job, and thus slowing down the system, we included some orchestration logic that ensures that the data population and data deletion jobs don't run at the same time.

The last challenge we will discuss is updates to the deletion logic, which may be required when changes are introduced to systems (such as a change in the DB schema). To support this type of change, we designed the solution such that plugins are loaded dynamically from a central repository upon execution of a deletion job. Thus, if a new version of a plugin is required, one simply needs to upload the new version to the repository and update the appropriate record in the System Registry. When the next deletion job is triggered, the new plugin logic will automatically be downloaded and run.

\section{Conclusion and future work}
\label{Conclusion}
We presented a reusable design for the automatic management and execution of requests for erasure of personal data in large-scale, evolving, data-intensive cloud-based environments. Besides saving time and preventing mistakes, this solution also makes it easy to track the progress of requests, see if/where things are stuck, and immediately remediate any issue to enable their timely completion. This is very important, especially in complex environments where data may be distributed over many different systems and datastores in many geographic locations. In addition, evidence of the deletion process execution is automatically documented and attached to the request, making it easy to retrieve in case of an inquiry or audit. 

Moreover, we demonstrated that this design is useful not only for compliance with data protection regulations but also for additional use cases such as adherence to business agreements between companies. A similar framework, with minor tweaks, could even be used to fulfill the GDPR's right to data portability (Article 20), by replacing the deletion commands with data retrieval commands.

There are a few more steps in the process that could be automated. These include the automatic creation of Jira tickets upon receiving erasure requests and automatic replies to data subjects whose requests are complete. Another area for future development is implementing more complex data retention policies that enable setting minimum and maximum retention periods, checking those against erasure requests, and enable adding future deletion tasks to be executed when the retention period passes.

In our current implementation it was assumed that all user data should be deleted, regardless of the purpose for which it was collected. It could however be extended to enable purpose-specific deletion by using additional functionality of the DPCM component. For example, a user could request that only data used for marketing be removed, whereas data used for product support remain stored in the system.

%
%
%
%

\end{document}